\documentclass[conference]{IEEEtran}
\IEEEoverridecommandlockouts
\usepackage{cite}
\usepackage{amsmath,amssymb,amsfonts}
\usepackage{algorithmic}
\usepackage{graphicx}
\usepackage{textcomp}
\usepackage{xcolor}
\def\BibTeX{{\rm B\kern-.05em{\sc i\kern-.025em b}\kern-.08em
    T\kern-.1667em\lower.7ex\hbox{E}\kern-.125emX}}
\begin{document}

\title{Benchmark Dataset for Timetable Optimization of Bus Routes in the City of New Delhi\\
}

\author{Anubhav Jain, Avdesh Kumar, Saumya Balodi, Pravesh Biyani\\
IIIT Delhi, India\\
{\tt\small \{anubhav15129, avdesh15135, saumya15172, praveshb\}@iiitd.ac.in}
}


\maketitle

\begin{abstract}
Public transport is one of the major forms of transportation in the world. This makes it vital to ensure that public transport is efficient. This research presents a novel real-time GPS bus transit data for over 500 routes of buses operating in New Delhi. The data can be used for modeling various timetable optimization tasks as well as in other domains such as traffic management, travel time estimation, etc. The paper also presents an approach to reduce the waiting time of Delhi buses by analyzing the traffic behavior and proposing a timetable. This algorithm serves as a benchmark for the dataset. The algorithm uses a constrained clustering algorithm for classification of trips. It further analyses the data statistically to provide a timetable which is efficient in learning the inter- and intra-month variations. 

\end{abstract}


\section{INTRODUCTION}
Public transport is one of the most popular means of transportation in various metropolitan cities across the globe. According to the Economic Survey of Delhi 2005-06, buses account for nearly 60\% of the total demand. While it is well understood that the public transport helps in combating air pollution and congestion caused due to single-occupancy vehicles, the usage of buses in Delhi and other cities in India has seen a nominal decline while the overall travel demand has simultaneously increased. \par
One of the main reasons for this decline in the city of Delhi (and other cities in India) is the lack of reliability of the bus routes. The timetable is often not made by the transit authorities. Moreover, it is often outdated soon due to the rapid change in infrastructure and the traffic conditions resulting in degradation in the reliability of buses. Finally, this decrease in reliability leads to unknown waiting times at the bus stops for the passengers.  Due to the absence of a timetable, most bus trips are operated in an ad-hoc fashion making it extremely difficult for the passengers to trust the public transport network leading to a decrease in passenger trips. \par
Interestingly, the various trips in a given bus-route still follow a certain pattern in a given day thanks to the pattern in the traffic conditions throughout the day. In other words, even when the transit operators do not follow an ``explicit" timetable, there is an ``implicit" timetable that is followed which is not completely random. The main goal of this work is to unearth this pattern and develop an operational timetable of the bus routes in the city of Delhi. The efficacy of this ``suggested" timetable is measured in terms of the average waiting time a passenger has to endure at various stops in the given route assuming she follows the timetable. This waiting time should ideally be lower than when the passenger does not follow the suggested timetable arrives at the same stop in a random fashion. \par 

The paper presents a novel database containing real-time data for over 500 routes of buses operating in Delhi. To arrive at the timetable for benchmarking the dataset, the paper samples two routes operated by the Delhi transport corporation. Out of the two routes, one is frequent, while the other operates at an average frequency of thirty minutes. The paper uses the collected GPS feed of the buses. 

The paper presents an approach which simplifies the problem statement to propose an efficient algorithm for defining the timetable based on GPS bus data collected over a period of two months.

\section{Related Work}
Researchers have been exploring the problem of reducing waiting time as well as reducing bunching in buses. Patnaik et al. \cite{patnaik2004estimation} applied data mining on data collected using Automatic Passenger Counters. The paper uses clustering to divide data points into different headways based on a decision tree. The method informs whether the existing headway would work for the buses. Wang et al. \cite{wang2017operating} used a sequential clustering method to ensure a fixed order for buses so that time division could be properly applied. They calculated the travel time by incorporating bus dwell time at stops which depends on the number of riders and also the road and intersection time depending on the traffic on the road at that time. Kornfeld et al. \cite{kornfeld2014optimizing} proposed an approach to minimize waiting time using two models which are based on the assumption of how people arrive at the bus stop. 

Yang et al. \cite{yang} proposed a method to optimize the timetable for the subway system by proposing an integer programming model to maximize the overlap time with the headway time. Wihartiko et al. \cite{Wihartiko} also proposed an integer programming model which uses a modified generic algorithm for timetabling of buses. Chakroborty et al.\cite{chakroborthy} and Deb et al. \cite{deb}, a mixed integer nonlinear programming model was proposed for timetabling of buses where the objective was to minimize the total waiting time of passengers. Saargunawathy et al. \cite{manogaran2017analysis} studied Open traffic platform which analysis traffic data linked to open street map. They used it to analyze the traffic conditions in Kuala Lumpur. They listed different roads and junctions with heavy traffic flow at different times. Sunil et al. \cite{gunjal2014dynamic} proposed a dynamic GPS based time-tabling algorithm for public transport that can predict the waiting time considering the real-time location of the user and the bus, and thus predicting estimated time of arrival.
 
The major issue with current research in the domain of timetable optimization for buses is that there is not a standard dataset which is being used by researchers. Various algorithms are applied on various datasets, which does not provide conclusive evidence of improvement in the state of the art. 

The proposed algorithm used for benchmarking the proposed dataset goes beyond the ones in literature in two ways. Firstly, literature does not take the simplicity of the algorithm into consideration. This is vital primarily as the IT Department of most transportation corporations are not educated enough to work with complex algorithms. The proposed algorithm closely revolves around the traffic behavior to provide the most optimal timetable. This algorithm is being deployed to update the timetable of over a hundred buses in New Delhi, India which would affect the lives of millions of daily bus users.

\begin{figure}[ht!]
  \centering
\includegraphics[width=\columnwidth]{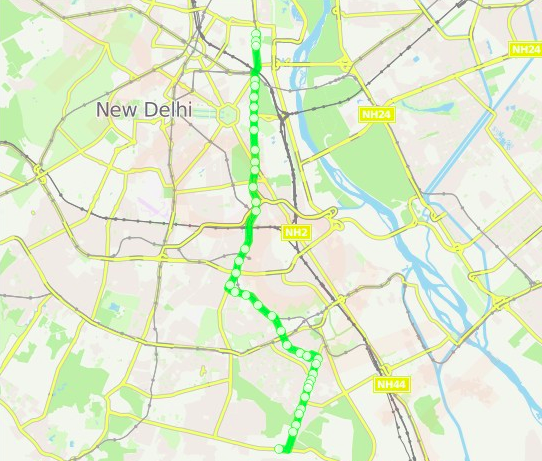}
\caption{Route for bus no. 425}
\label{fig:425_Route}
\end{figure}

\begin{figure}[ht!]
  \centering
\includegraphics[width=\columnwidth]{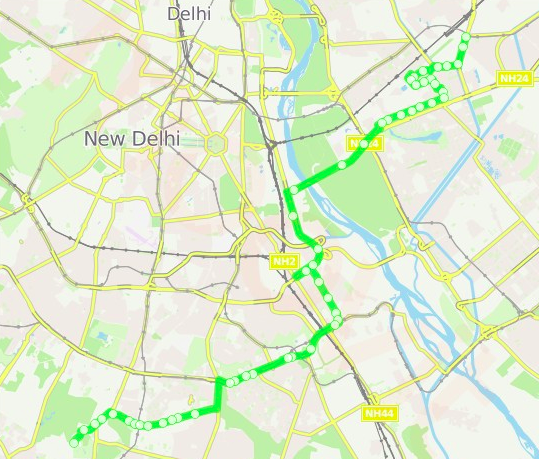}
 \caption{Route for bus no. 534}
\label{fig:534_Route}
\end{figure}

\begin{figure}[ht!]
  \centering
\includegraphics[width=\columnwidth]{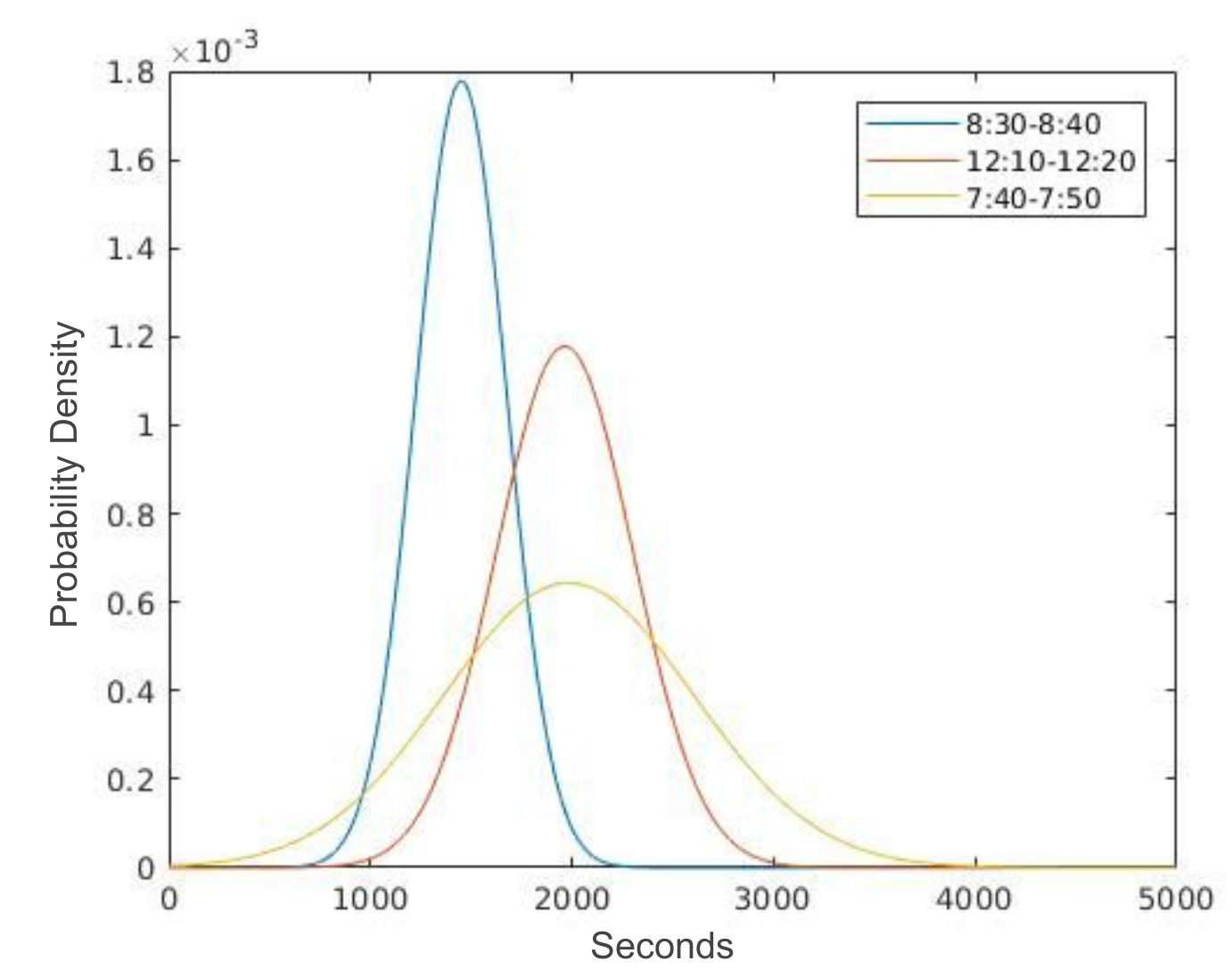}
  \caption{Probability distribution of the arrival time at stop 17 for route no. 425 at different times of the day.}
\label{fig:425_pdf}
\end{figure}

\section{Proposed Database}

The proposed bus transit database \footnote{https://bus-data.github.io/} contains (i) Static and (ii) Dynamic data for buses operating in Delhi. The dynamic data for three months has been stored separately which can be used for applications which require previous data.  This is the first of its kind database which provides access to real-time transit data. This dataset would provide means for standardization of timetable optimization algorithms which was not possible earlier as researchers were using classified datasets which weren't available publicly. This information can easily be used for building real-time applications which span from building timetable optimization algorithms, which has been benchmarked in this paper along with various other applications such as live GPS tracked monitoring of traffic, tracking anomalies in live traffic data for security purposes, travel time estimation and optimization. As the live GPS data is sampled at every 10s for all the routes, it provides substantial information about different roads on in the city and the expected travel speed possible on those routes.   

For creating the database, the paper has pre-processed the data using static information about the bus stops for that particular route. Through the initial step of pre-processing, the data was divided into relevant bus stops which were considered to be nodes. To do so, every raw data point which lied within a 50-meter radius of the coordinate of a bus stop was mapped to the stop. The 50 meter threshold is chosen keeping in mind situations where the bus traveling at a high speed might not store a data point close to the bus stop. By doing so, the raw data was narrowed down to the stop based data where all the nodes were the stops. This brought uniformity for further processing of the data. The bus routes are represented in figure \ref{fig:425_Route} and figure \ref{fig:534_Route} for routes 425 and 534 respectively. The route from the first stop to the last is considered as the up route, and the opposite direction is referred to as the down route.

 \subsection{Dataset Statistics}
 
 The dataset captures the following two categories, mentioned below. 
 
 \subsubsection{Dynamic Data} 
 
 The data for all 543 routes and 3464 stops are provided real-time. The data was in the form of Protocol Buffers and consists of information about all the buses operating at that time. The data provided has a frequency of 10s, where each data packet contains the date, time, latitude and longitude values for each bus route along with bus-specific details like number plate, route number, direction of route (up/ down). This can easily be accessed for real-time applications and can be stored and used as statistical data. The data tracks all the buses throughout the time they are active.
 
 \subsubsection{Static Data}
  The static data provides information about the routes, stops, trips and stops times for all the buses which are being modeled dynamically. The information contains specific information to use dynamic data and extract the required information. The dynamic data is coded using the information provided in the static data. This compresses the dynamic data packets and makes it more efficient for real-time applications.

 To use this data for optimizing timetable, the dynamic data has been stored as for over three months in a database which can be directly used for the application of timetable optimization algorithms. The data has been decoded and is provided in a directly usable format. This has been done for routes 425 and 534. These datasets have also been benchmarks using the proposed approach.

\section{Proposed Approach}

Current bus timetables are formulated without considering the traffic behavior based on time of the day and the day of the week. It has been observed from the data that these are significant components which affect the arrival time of the bus at a particular stop which needs to be carefully scrutinized to provide a more efficient timetable. 


The paper provides benchmarks for two routes- 425 and 534. These buses were specifically chosen as they differ quite significantly in their frequency and reliability. To model an efficient timetable, the paper looks at these two buses where 534 is extremely frequent and has an average waiting time of around 5 minutes while buses on route 425 are more unreliable and have an average waiting time close to 22 minutes. 




The variables are defined as:

\begin{itemize}
    \item Total bus stops $N$ are present on the route. 
    \item The data has been taken for $d$ days. 
    \item $i$ represents the trip variable, $\in(0,x^k)$. 
    \item $j$ denotes the stop variable, $\in(0,N)$.
    \item $k$ represents the day variable $\in(0,d)$. 
    \item $x^k$ represents the number of trips on the $k^{th}$ day. 
    \item $t^k_{i,j}$ represents the arrival time of the bus at the $j^{th}$ stop for the $i^{th}$ trip on the $k^{th}$ day.
    
\end{itemize}

To reduce the effect of noise in the data, the algorithm samples the data at every $3^{rd}$ stop. The algorithm can be summarized in two stages, definition of the starting time at the first stop and calculation of arrival time for the following stops based on the first stop.

The optimization problem finds the most optimal $\widehat t_{i,j}$ such that, 

\begin{equation}
    \widehat t_{i,j} = \min \quad Var[t^k_{i,j}]
\end{equation}



\subsection{Defining the starting time}

As the timetable depends on the departure time of the bus from the first stop, the algorithm uses K-means clustering \cite{hartigan1979algorithm} for this step. Unlike the standard clustering algorithm, which minimizes the inter-class variance, this approach minimizes the variances while keeping a minimum distance between the clusters. The minimum distance between two clusters is equivalent to the standard frequency of the buses as suggested by the Transportation Department. 

Each of the set of data points in each cluster created using this approach is represented by $C^{(n)}$ at iteration $n$ which contains $M^{(n)}$ data points. The centroid of the $l^{th}$ set of cluster is denoted by $\mu_l^{(n)}$ at iteration $n$ and the total number of clusters are $c$. Each of these cluster sets can be mathematically represented using equation \ref{eq:clustering}, where the new data point being classified is $tp$.  

\begin{equation}
    C_{l}^{(n)} = {\{tp: ||tp - \mu_{l}^{(n)}||^{2} < ||tp - \mu_{m}^{(n)}||^{2}, \forall m; 1<m<c }\}
    \label{eq:clustering}
\end{equation}

The definition of new clusters is conditioned on equation \ref{eq:mean_cond}, where $T_1$ is the frequency of the bus.  

\begin{equation}
    {\{||\mu_{l}^{(n)} - \mu_{m}^{(n)}|| > T_1, \forall l,m; \space \space 1<l,m<c }\}
    \label{eq:mean_cond}
\end{equation}

The mean is updated using the equation \ref{eq:clustering_mean}. 

\begin{equation}
    \mu_{l}^{(n+1)} = \frac{ \sum_{ t^{k,(n)}_{i,j} \in C_{l}^{(n)}}^{}  t^{(n),k}_{i,j}}{|C_{l}^{(n)}|} = \frac{\mu_{l}^{(n)}*M^{(n)}+ tp^{(n+1)}}{M^{(n)}+1}
    \label{eq:clustering_mean}
\end{equation}

In the first step, the average departure time for a bus is calculated. While considering each new bus, the nearest mean was searched, and if the difference between the departure time and the nearest mean is less than the threshold $T_1$, then the mean of that cluster was updated by including this data point as well. Otherwise, this would be considered as a new cluster. Finally, only if the total number of points in a cluster is greater than a threshold $T_2$, the starting time is considered valid. Otherwise, the data point is considered to be an outlier case and is not considered for further calculations. Due to noise in the data and inconsistency in the starting time of the first bus, the paper grid searches for the best threshold to find the appropriate starting time which reduces the waiting time as well as the bunching of buses. The threshold $T_2$ is taken as ten days which is one-third of the total number of days for which the data is used. 







\subsection{Adjacent Stops}

For every subsequent bus stop, the average time to reach that bus stop is calculated from the first node for every 15-minute interval. The timetable for these nodes is defined as the start time plus the average time taken to reach that stop (in the particular time range). The optimal time for every adjacent stop can be written as, 

\begin{equation}
    \widehat t_{i,j} = \widehat t_{i,1} + \frac{\sum_{k=1}^d (t^k_{i,j} - \widehat t_{i,1})}{d} \quad \forall j \in(2,N)
\end{equation}

Figure \ref{fig:425_pdf} shows that the arrival time of the bus at each stop is in the form of a Gaussian distribution which has higher variance as the day progresses. By taking the mean in the above approach, the algorithm minimizes the waiting time, which is defined as the variance of this distribution — thus giving the most optimal timetable. 

The data is first divided into 15-minute time slots for more accurate calculation and estimation of traffic behavior. This dynamic time-based timetable takes into consideration the regular changes which come in the traffic behavior every 15 minutes.

\subsection{Calculating Waiting Time}

The pre-timetable waiting time (preWT) is defined as the expected time a person will wait if he/she arrives at a bus stop at any random time. If the next bus arrives after $N_0$ minutes from the previous bus then the expected waiting time for the next bus at that stop will be:

\begin{equation}
    preWT = \sum_{n=1}^{N_0} n/N_0 = (N_0+1)/2
\end{equation}


The algorithm averages this over all the trips for a bus route to get the mean waiting time for different stops. 


For calculating the post-timetable waiting time, equation \ref{eq11} has been used. Where the mentioned instantaneous post-timetable waiting time (ipostWT(i,j,k)) has been averaged over all the days/ trips for various experiments. 

\begin{equation}
    ipostWT(i, j, k) =  \min\limits_{x} (t^k_{i,j} - \widehat t_{x,j}) \quad s.t.  \quad t^k_{i,j} > \widehat t_{x,j} 
    \label{eq11}
\end{equation}




\begin{figure}[ht!]
  \centering
\includegraphics[width=1.12\columnwidth]{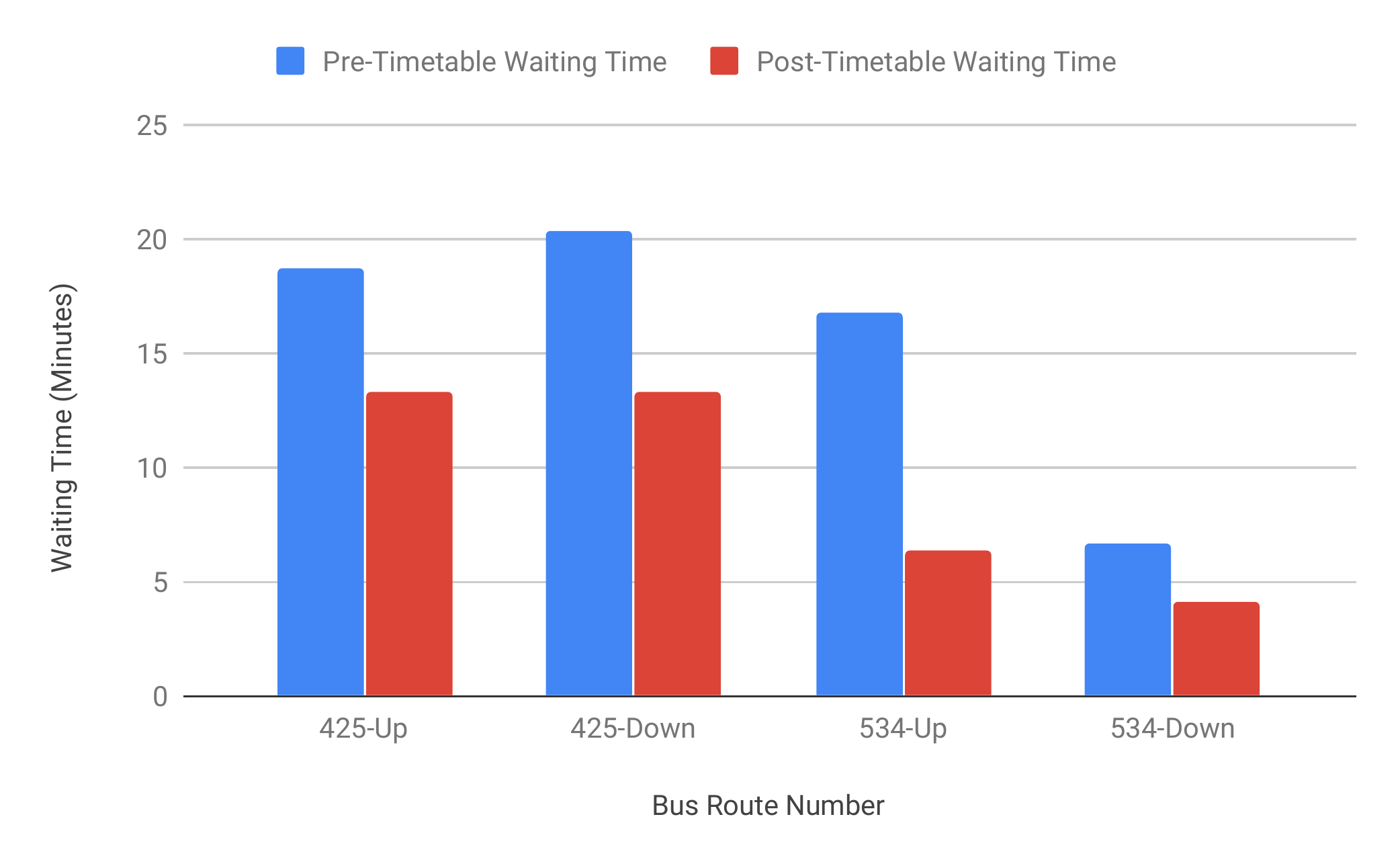}
 \caption{Comparison of Waiting Time at the $1^{st}$ Node for all the Buses.}
\label{fig:first}
\end{figure}

\begin{figure}[ht!]
  \centering
\includegraphics[width=1.12\columnwidth]{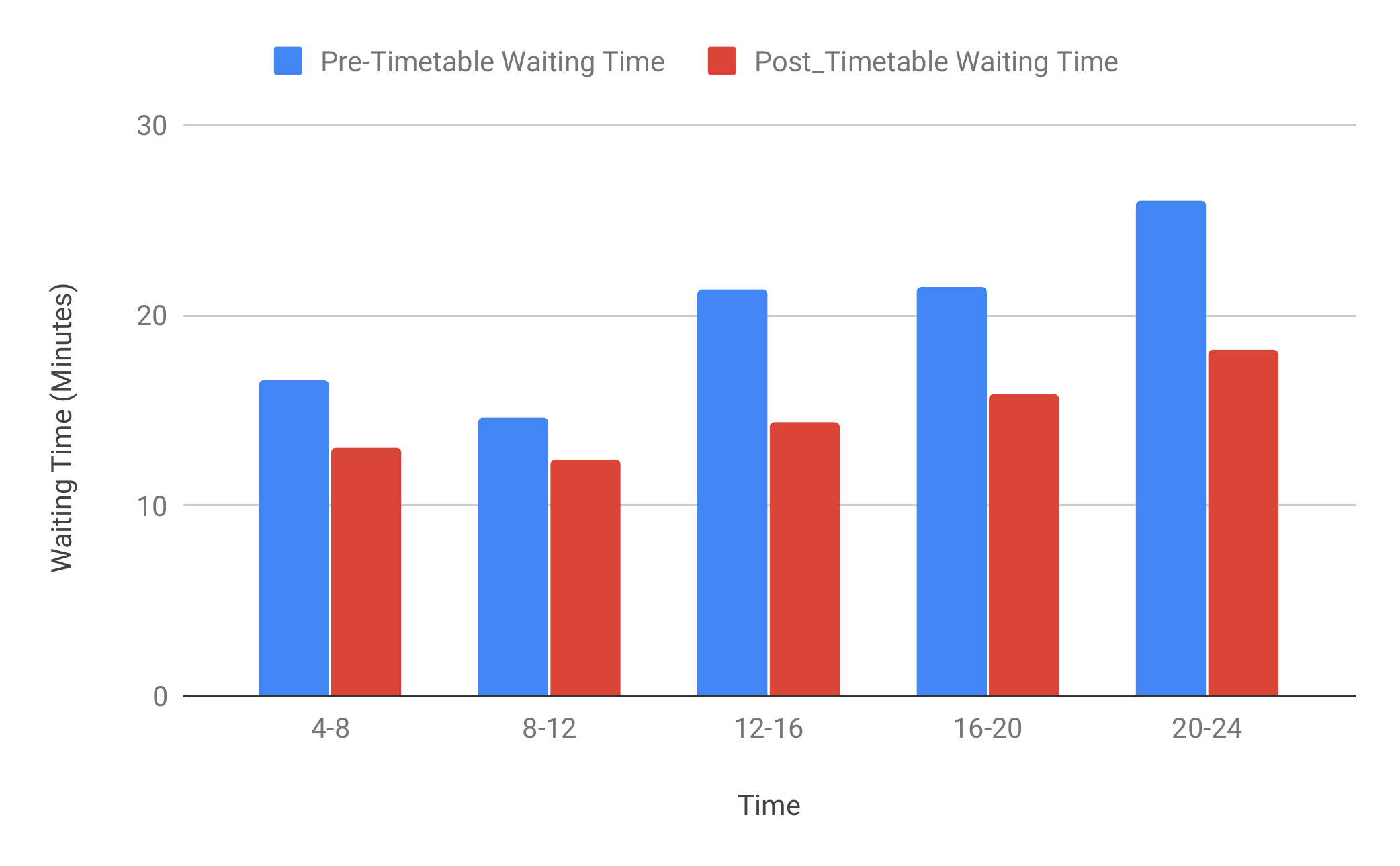}
 \caption{Comparison of waiting time for bus no. 425 Down Route before and after the timetable when training and testing on alternate days.}
\label{fig:425_down}
\end{figure}

\begin{figure}[ht!]
  \centering
\includegraphics[width=1.12\columnwidth]{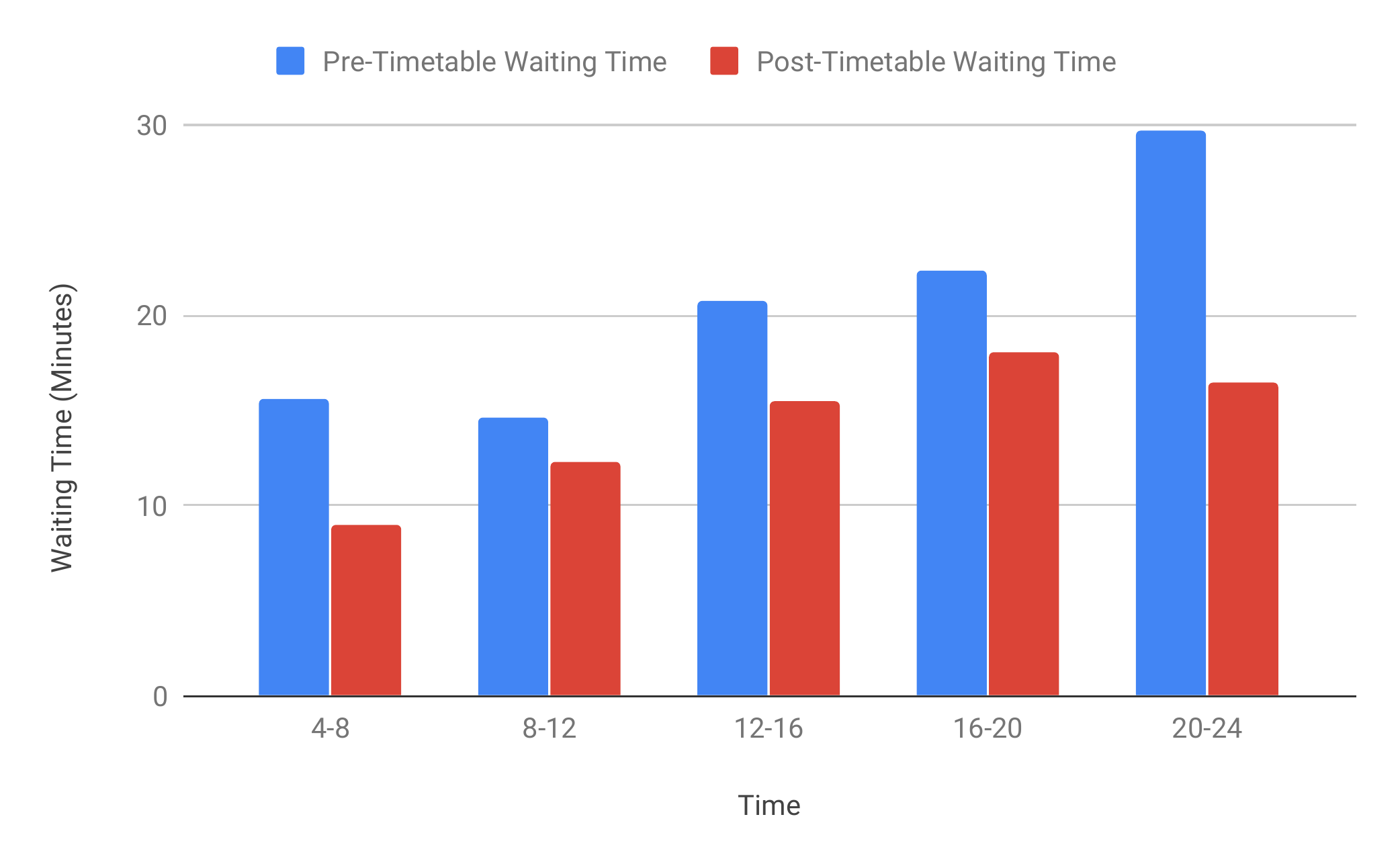}
 \caption{Comparison of waiting time for bus no. 425 Up Route before and after the timetable when training and testing on alternate days.}
\label{fig:425_up}
\end{figure}

\begin{figure}[t]
  \centering
\includegraphics[width=1.12\linewidth]{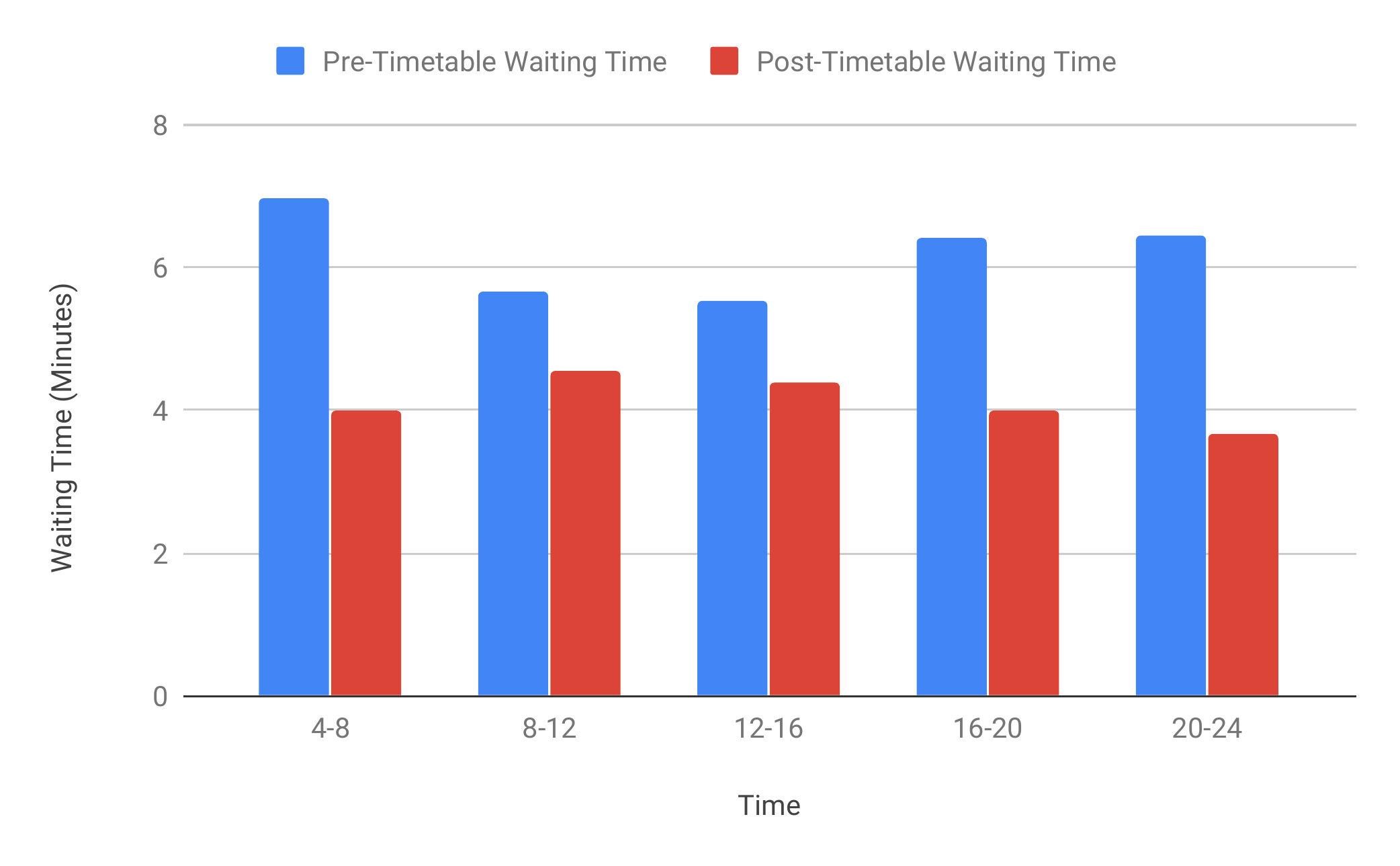}
 \caption{Comparison of waiting time for bus no. 534 down route when training and testing on alternate days.}
\label{fig:534_down}
\end{figure}

\begin{figure}[t]
  \centering
\includegraphics[width=1.12\linewidth]{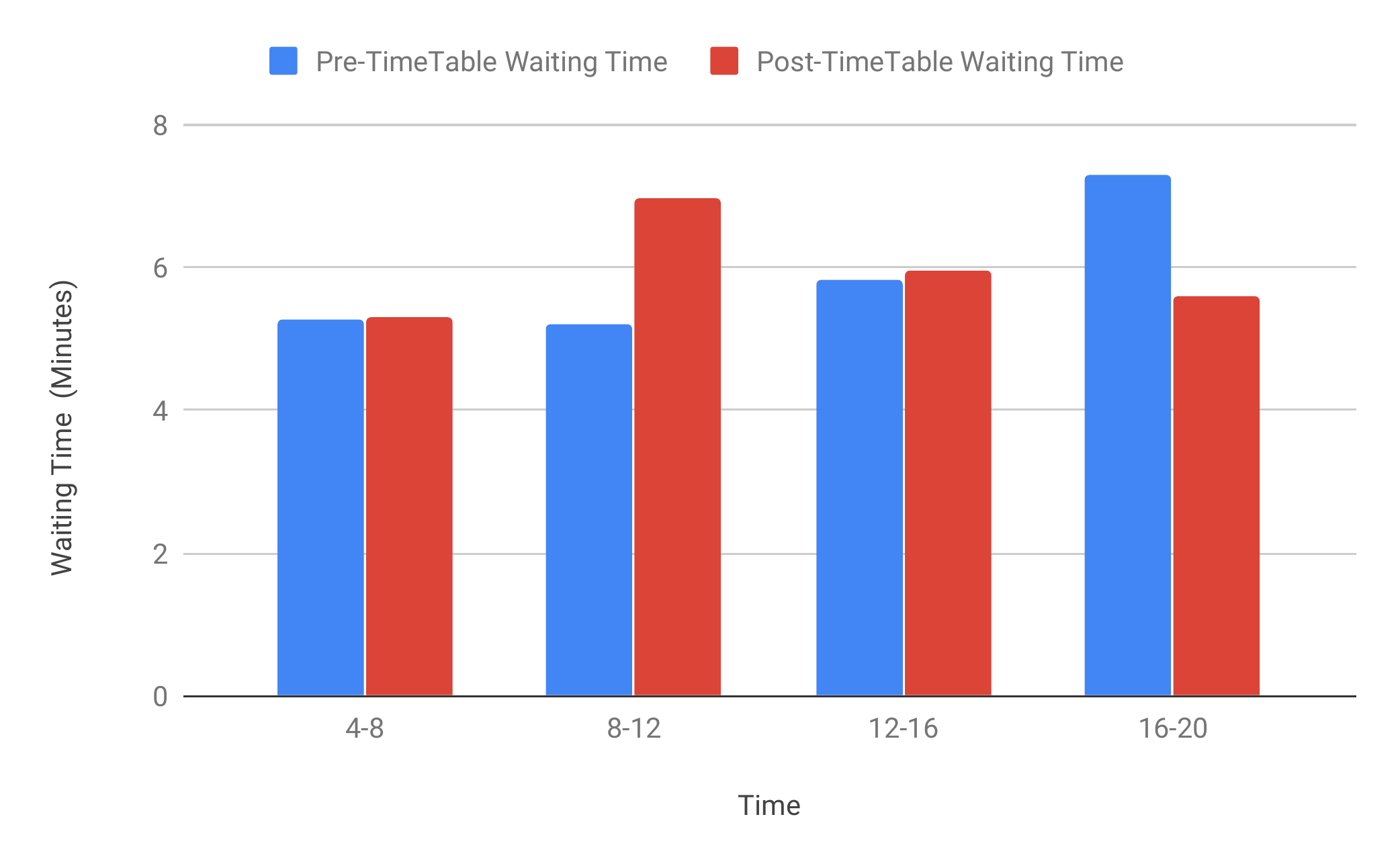}
  \caption{Comparison of waiting time for bus no. 534 up route when training and testing on alternate days.}
\label{fig:534_up}
\end{figure}


\begin{figure}[ht!]
  \centering
\includegraphics[width=1.12\columnwidth]{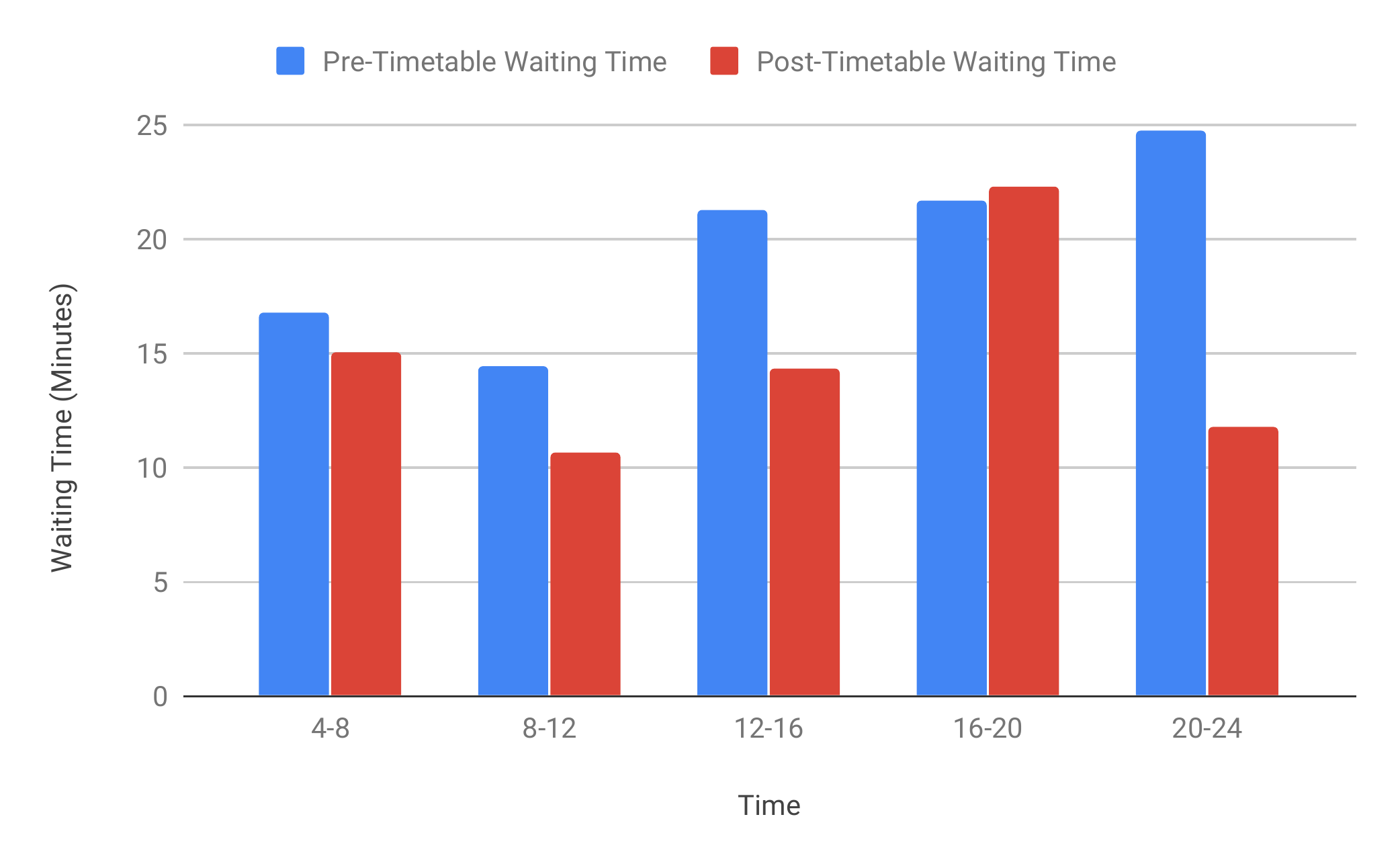}
 \caption{Comparison of waiting time for bus no. 425 Down Route before and after the timetable when testing on second month data.}
\label{fig:425_down-nov}
\end{figure}

\begin{figure}[ht!]
  \centering
\includegraphics[width=\columnwidth]{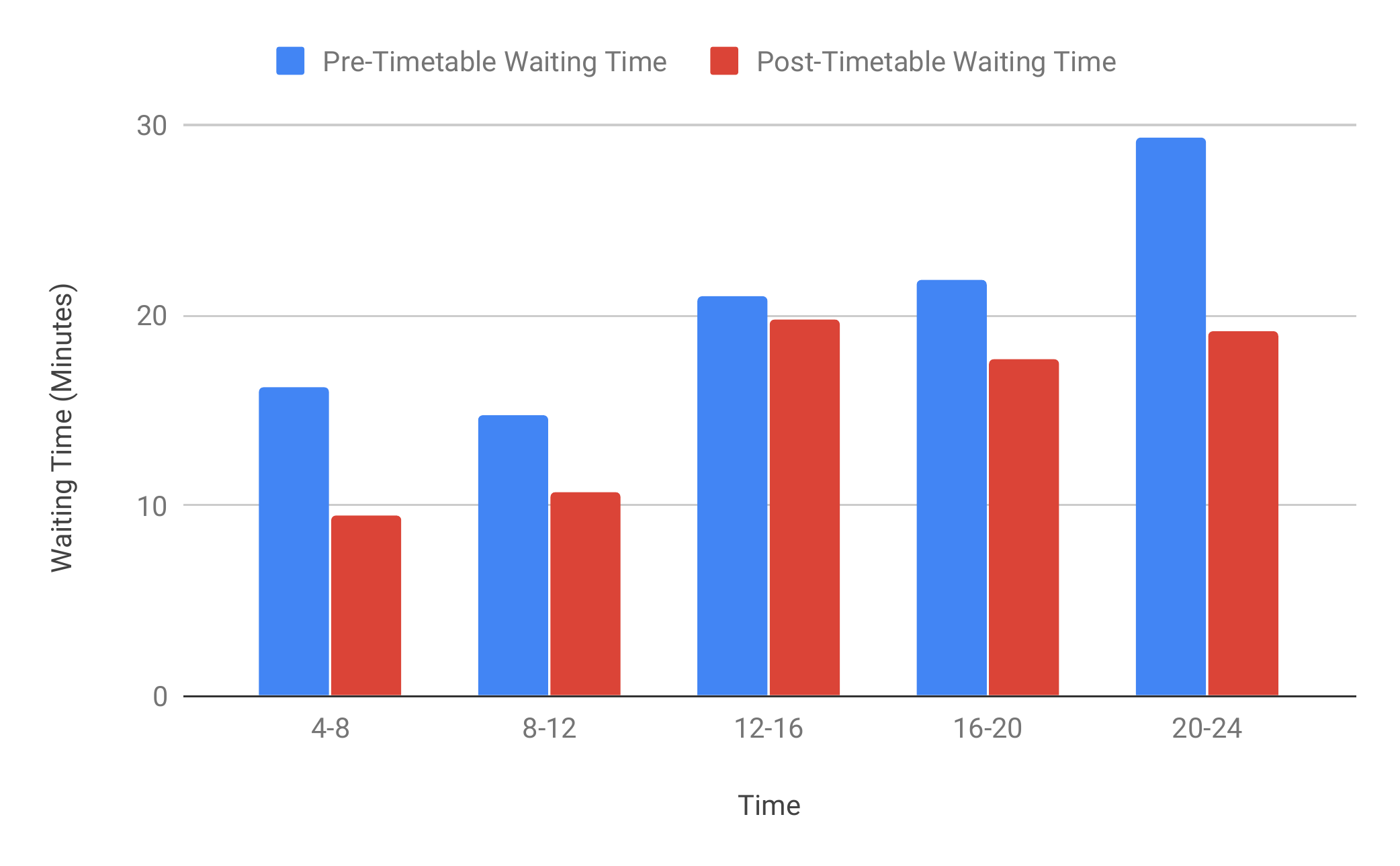}
 \caption{Comparison of waiting time for bus no. 425 Up Route before and after the timetable when testing on second month data.}
\label{fig:425_up-nov}
\end{figure}

\begin{figure}[t]
  \centering
\includegraphics[width=\linewidth]{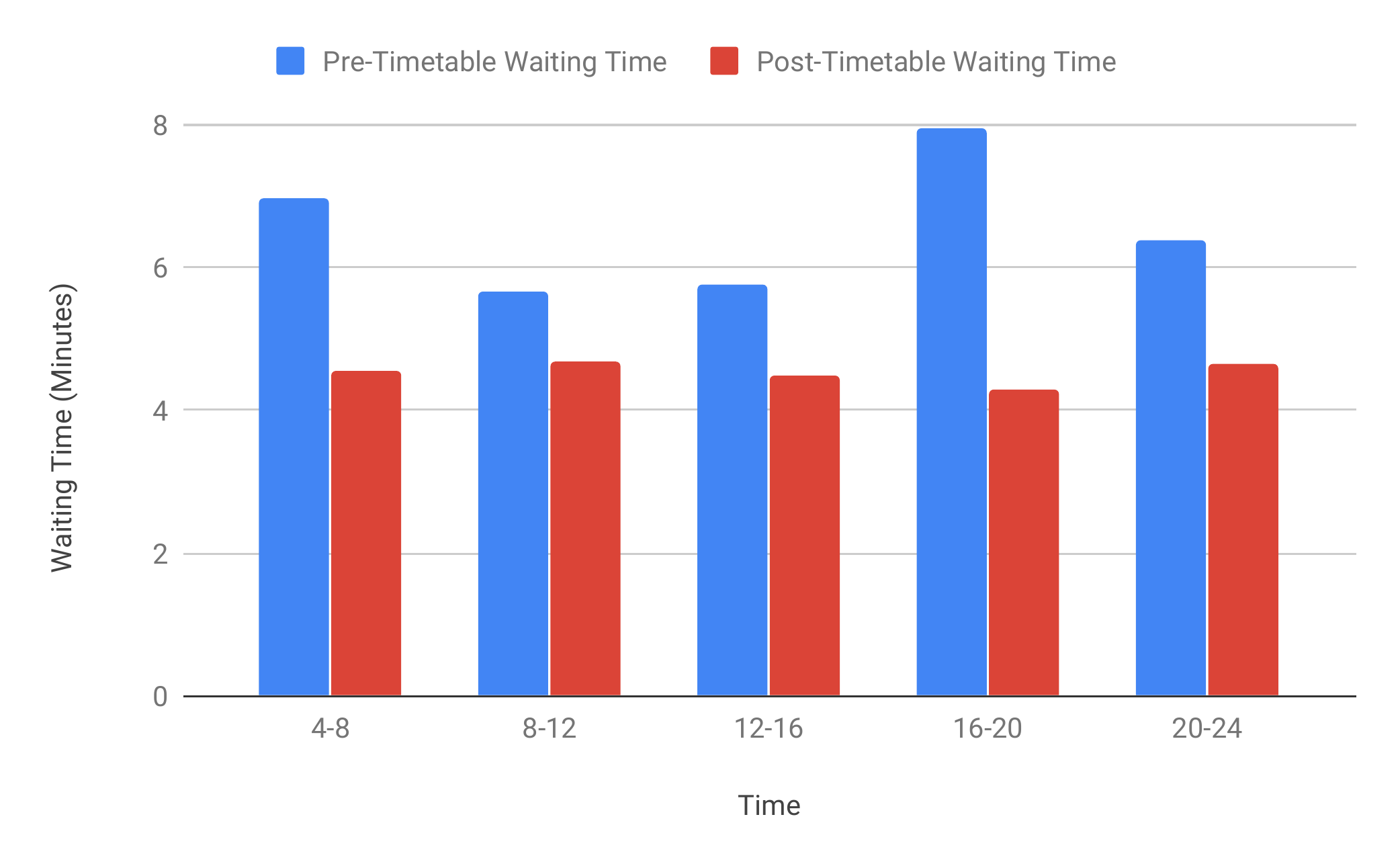}
 \caption{Comparison of waiting time for bus no. 534 down route when testing on second month data.}
\label{fig:534_down-nov}
\end{figure}

\begin{figure}[t]
  \centering
\includegraphics[width=\linewidth]{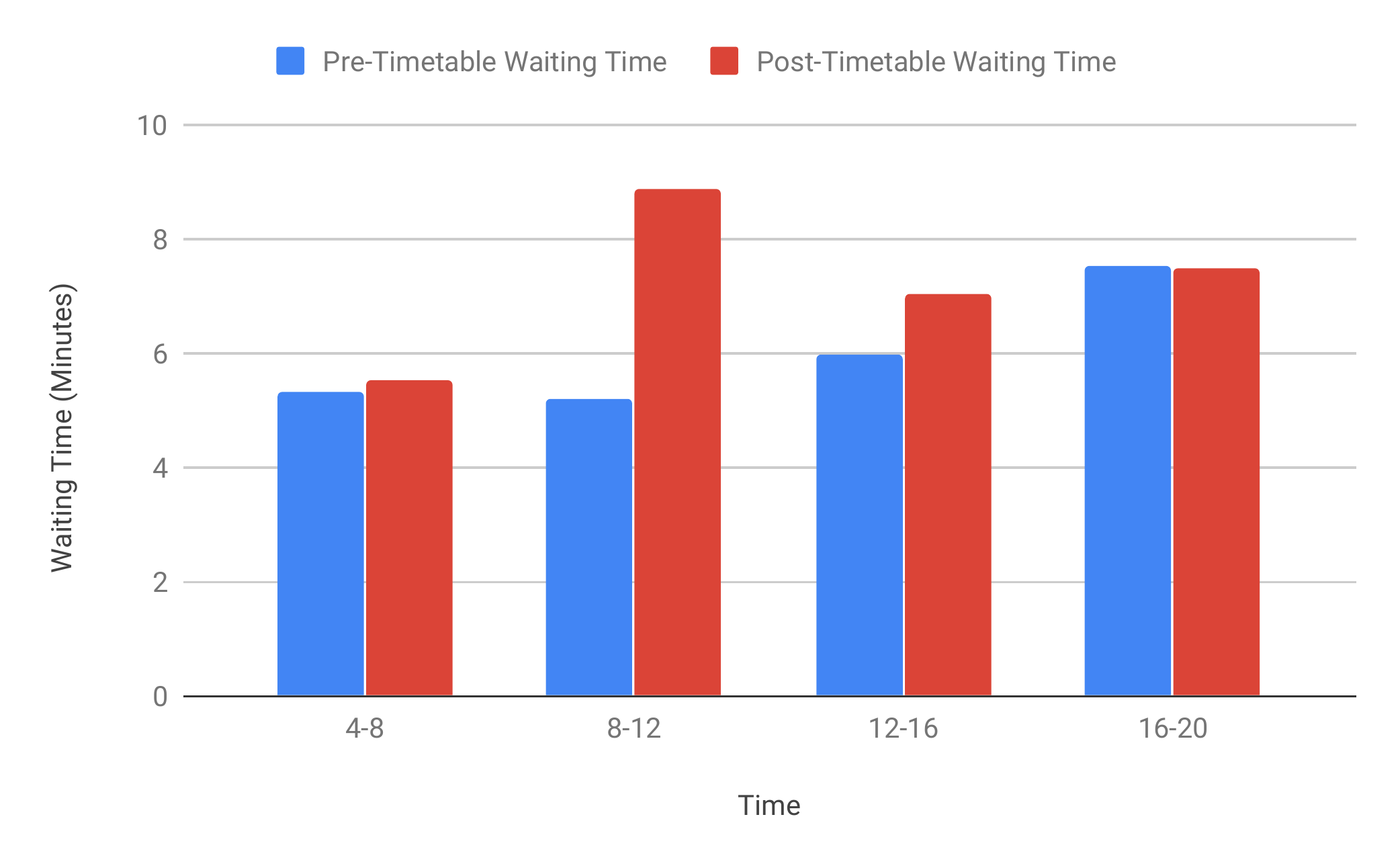}
  \caption{Comparison of waiting time for bus no. 534 up route when testing on second month data.}
\label{fig:534_up-nov}
\end{figure}

\section{Results}

The paper proposes an approach to create a timetable for the two buses using a relatively direct procedure which resulted in a significant decrease in the waiting time which has been applied to the two bus routes - 425 and 534. It has been applied separately to the up and down routes of the buses. 

The following experiments have been performed to show the efficiency of the algorithm. 
\begin{itemize}
    \item The first experiment shows the performance of using K-means clustering for classification of the data into trips. This experiment is specific to the first node, i.e., the starting point of the bus. The algorithm has been trained on the first-month data, and it has been tested on second-month data.
    \item To judge the model for the formation of the timetable of the remaining stops as well as learning intra-month variations, the model was trained and tested on alternate days of the first and second month. The data was trained on all the odd days of the two months, and it was tested on even days.
    \item For learning inter-month variations in a data which already contains high randomness, the algorithm was trained on first-month data and tested on the second-month data. 
\end{itemize}

The results have been presented in the form of bar graphs showcasing the difference brought by the introduction of the proposed timetable. The impact of clustering for the initial node, by following the first protocol, has been presented in figure \ref{fig:first}. This shows that there a significant amount of randomness in the starting time of these buses, as the current set timetable is not followed. The proposed approach has been able to reduce this randomness of the starting time. If this is strictly followed it would yield much more promising results for the subsequent stops as well. As the resultant increase in waiting time of the following bus stops is the addition of the randomness in the $1^{st}$ stop along with the randomness in the traffic behavior. The first can be controlled by proper implementation of the timetable by public/ private transportation departments. 


For the second experiment, the paper looks at the formation of a timetable by looking at the inter-month variations in the data. In doing so, the model shows its efficiency in learning the randomness which is present within the constraints of the same month. Figure \ref{fig:534_down} shows a comparison of the waiting time without any timetable with the waiting time post the new timetable for bus no. 534. Figure \ref{fig:534_up} shows the graph for the the up-route while figure \ref{fig:534_down} shows the graph for the down-route. Both the graphs compare the waiting time at different time intervals. 

Lastly, the model has been tested on a completely unseen data after being trained on the first-month data. The results upon being tested on the second-month data have been presented in figure \ref{fig:425_down-nov} and figure \ref{fig:425_up-nov} for the up and down route of bus number 425 respectively. 

The impact of a timetable is more in the case of bus route 425 due to higher irregularity. Due to the high frequency of bus number 534 introducing large changes is not possible. During the calculation of final waiting time, the randomness in the starting time due to human constraints have been taken into consideration. As the final waiting time is calculated using the same probabilistic distribution of the starting stop. Thus, showing that even if the timetable is not followed closely, it would still impact the waiting time considerably. These situations make this essential for real-world applications where human constraints are significant.



\section{Conclusion and Future Work}

The paper proposes a novel and first of its kind freely available benchmark dataset for analyzing real-time bus transit. The data captures real-time variations in traffic, and moreover, it can be used for standardization of timetable optimization algorithms, which are currently applied on a variety of datasets which aren't publicly available. For benchmarking the dataset, the paper presents an approach to propose a timetable taking into consideration traffic variations across time. The paper shows the efficiency of this algorithm in reducing the waiting time of passengers on Delhi bus routes. The algorithm is applied to the Delhi Transportation Corporation buses number 425 and 534. 

While this approach looks at the efficiency with Delhi based buses, it would be interesting to see the implementation of buses from different cities. The paper assumes that the traffic behavior does not vary to a large extent across days. This is another factor which can be added while defining the timetable. Another exciting area of research would be incorporating a limit on the number of passengers. The timetable can be established to keep a check on such factors which influence passenger satisfaction.

\bibliographystyle{ieeetr}
\bibliography{bibtex}




\end{document}